\documentclass[12pt]{iopart}

\newtheorem{thm}{Theorem}

 \newtheorem{lem}{Lemma}
 \newtheorem{prop}{Proposition}

\begin{document}

\title[N. Li]{A new 3-component Novikov hierarchy}

\author{Nianhua Li}
\address{School of Mathematical Sciences, Huaqiao University, Quanzhou, 362021, P R China}
\ead{linianh@hqu.edu.cn}

\begin{abstract}
We study the bi-Hamiltonian structure of the hierarchy of a 3-component Novikov system. We show that Hamiltonian functionals of the 3-Novikov hierarchy in negative direction are local, and in both directions are homogenous. We construct a reciprocal transformation to connect the 3-Novikov system to a reduction of the first negative flow in a modified  Yajima-Oikawa hierarchy, which is shown to pass the standard Painlev\'{e} test. Besides we discuss bi-Hamiltonian structures of the 3-Novikov hierarchy under the reciprocal transformation. Moreover, we consider a limit for the 3-Novikov system.
\end{abstract}

\section{Introduction}
The Camassa-Holm (CH) equation
\begin{equation}\label{ch}
m_t+um_x+2u_xm=0,\quad m=u-u_{xx},
\end{equation} has attracted much attention since it is derived as the governing equation for dispersive shallow-water motion in 1993 \cite{Holm}. It is remarkable that the CH equation has  peakon solutions which are interesting in general analysis of PDEs \cite{Lun}. The CH equation is integrable from the point of view of Lax pair and bi-Hamiltonian structure \cite{Holm,Hyman}. It is linked to the negative  KdV equation by a reciprocal transformation \cite{Fuch,Lenells,Mckean}. In Ref. \cite{Qu} and its references, many other algebraic and geometric properties of the CH equation are introduced.

By applying asymptotic integrability method to a family of
third order dispersive PDE, Degasperis and Procesi \cite{Degas} found another equation possessing peakon solutions
\begin{equation}
m_t+um_x+3u_xm=0,\quad m=u-u_{xx}.
\end{equation}The DP equation  has a Lax pair and a bi-Hamiltonian structure  \cite{DHH}. An infinite sequence of conservation laws for the equation are also obtained. Besides a reciprocal transformation is constructed to connect it with a negative flow in the Kaup-Kupershmidt hierarchy.
Hereafter, many other equations of CH type were proposed and studied. For example, the Novikov equation, the modified CH equation, a 2-component CH equation and the Geng-Xue equation  (see e.g. \cite{Novikov,Wang,Geng,Hone,Olver,Chen}).

Recently, Geng and Xue \cite{Xue} presented a 3-component CH type hierarchy by consider the following $3\times 3$ matrix spectral problem
\begin{equation}\label{lax0}
\varphi_x=\left(
            \begin{array}{ccc}
              0 & 1 & 0 \\
              1+\lambda u & 0 & v \\
              \lambda w & 0 & 0\\
            \end{array}
          \right)
\varphi.
\end{equation} The spectral problem (\ref{lax0}) may reduce to that of the CH equation, the DP equation, the Novikov equation and the Geng-Xue equation.  The corresponding hierarchy was derived by choosing the trivial flow as $(u,v,w)^{T}_t=(u,v,w)^{T}_x$. The first negative flow in the hierarchy reads
\begin{eqnarray}\label{gengxue3}
&&u_{t}=-v p_{x}+u_{x}q+\frac{3}{2}u q_{x}-\frac{3}{2} u(p_{x}r_{x}-pr),\nonumber\\
&&v_{t}=2vq_{x}+v_{x}q ,\nonumber\\
&&w_{t}=v r_{x}+w_{x} q+\frac{3}{2}w q_{x}+\frac{3}{2} w(p_{x}r_{x}-pr),\label{h1}\\
&&u=p-p_{xx},\quad \quad \quad \quad \quad \quad w=r_{xx}-r,\nonumber\\
&&v=\frac{1}{2}(q_{xx}-4q+p_{xx}r_{x}-r_{xx}p_{x}+3p_{x}r-3p r_{x}).\nonumber
\end{eqnarray}
This system can be reduced to the CH equation as $p=r=0$. It admits a bi-Hamiltonian structure and an infinite sequence of conserved quantities \cite{Xue,Li}.  However, it is hard to construct some exact solutions for this system.

Subsequently, by considering reductions of a 4-component CH type system, we proposed another 3-component CH type system
\begin{eqnarray} \label{3eqq}
\begin{array}{l}m_{1t}+u_2gm_{1x}-m_{3}(u_{2x}f-u_2g)-m_1(3u_2f-m_3u_2)=0,\\
m_{2t}+u_2gm_{2x}+m_{2}(3u_{2x}g+m_3u_2)=0,\\
m_{3t}+u_2gm_{3x}-m_3(2u_2f+u_{2x}g-m_3u_2)=0, \\
m_{i}=u_{i}-u_{ixx}, \ i=1..3, \quad f=u_{3}-u_{1x},\quad g=u_{1}-u_{3x}, \end{array}
\end{eqnarray}
associated with the spectral problem \cite{Popowicz}
\begin{equation}\label{thr2}
\phi_{x}=\left(
    \begin{array}{ccc}
      0 & 0 & 1 \\
       \lambda m_{1}& 0 & \lambda m_{3}\\
      1 &  \lambda m_{2}& 0 \\
    \end{array}
  \right)\phi.
\end{equation} It is shown to possess a bi-Hamiltonian structure and infinitely many conserved quantities. The system (\ref{3eqq}) is found to connect with a negative generalized MKdV system (a modified Yajima-Oikawa (mYJ) system\cite{Yajima,Liu}) via a reciprocal transformation, and the associated system is shown to pass the standard Painlev\'e test of WTC \cite{Li2}.

In this paper,  we will study a new 3-Novikov hierarchy
associated with the following spectral problem
\begin{equation}\label{lax1}
\varphi_x=U
\varphi,\quad U=\left(
            \begin{array}{ccc}
              0 & 1 & 0 \\
              1+\lambda u^2 & 0 & v \\
              \lambda w & 0 & 0\\
            \end{array}
          \right), \quad \varphi=\left(
  \begin{array}{c}
    \varphi_1 \\
    \varphi_2 \\
    \varphi_3 \\
  \end{array}
\right),
\end{equation}which is obtained by replacing $u$ in (\ref{lax0}) with $u^2$ for convenience. The new hierarchy is different from the hierarchy found by Geng and Xue, because we take the trivial flow as $(u,v,w)^{T}_t=(0,v,-w)^{T}$.
The first typical member in the 3-Novikov hierarchy is the 3-Novikov system
\begin{eqnarray}\label{3ch}
\begin{array}{l}u_t+(upr)_x=0,\\
v_t+3vp_xr+v_xpr+u^2p=0,\\
w_t+3wpr_x+w_xpr-u^2r=0,\\
v=p-p_{xx},\quad \quad \quad w=r-r_{xx}.\end{array}
\end{eqnarray}
It can be reduced to the DP equation, the Novikov equation and the Geng-Xue equation as $u=0, r=1$, as $u=0, p=r$ and as $u=0$ respectively. We will construct infinitely many conserved quantities and study the bi-Hamiltonian structure of the 3-Novikov hierarchy, and construct a reciprocal transformation for the system (\ref{3ch}).

The outline of this paper is as follows. In Section 2, we construct infinitely many conserved quantities for the 3-Novikov equation with the aid of the spectral problem (\ref{lax1}). We also analyze the homogeneous and local properties of the Hamiltonian functionals in the 3-Novikov hierarchy. In Section 3, we find the relationship between the two systems (\ref{3eqq}) and (\ref{3ch}). In Section 4, we construct a reciprocal transformation to connect the 3-Novikov system with the first negative flow in a mYJ hierarchy, and analyse the bi-Hamiltonian structure under this transformation. In Section 5, we present a limit of the 3-Novikov system. 
\section{Conserved quantities and bi-Hamiltonian structure of the 3-Novikov hierarchy}

\subsection{Conserved quantities}
The 3-Novikov system (\ref{3ch}) arises as the compatibility condition for the linear system
\begin{equation}\label{Laxpair}
\varphi_x=U
\varphi,\quad \quad \quad \varphi_t=V
\varphi,
\end{equation}where
\begin{equation*}
V=\left(
            \begin{array}{ccc}
              \frac{1}{3\lambda}+pr_x & -pr & \frac{p}{\lambda} \\
              p_xr_x-\lambda u^2pr & \frac{1}{3\lambda}-p_xr & \frac{p_x}{\lambda}-vpr \\
              -\lambda wpr-r_x & r & p_xr-pr_x-\frac{2}{3\lambda}\\
            \end{array}
          \right).
\end{equation*}
With the aid of the Lax pair (\ref{Laxpair}), infinitely many conserved quantities or conservation laws for the 3-Novikov system can be constructed. For example, setting $\rho=({\rm ln}\varphi_3)_x$ and expanding it in powers of $\lambda$, as pointed out in  \cite{Xue}, one may able to obtain an infinite sequence of conserved densities for (\ref{3ch}) from coefficients of $\rho$ by solving
\begin{equation}\label{conserved}
(\partial+\rho)[(\frac{\rho}{w})_x+\frac{\rho^2}{w}]-(1+\lambda u^2)\frac{\rho}{w}-\lambda v=0.
\end{equation}
However, it is not easy to solve (\ref{conserved}) and the expansion of $\rho$ in  \cite{Xue} can be generalized. Therefore we will consider a better formulation for computations and get more exact conserved quantities, which may be useful to generalize flows of the 3-Novikov hierarchy and to construct reciprocal transformations.

Let $a=\frac{\varphi_1}{\varphi_3}, b=\frac{\varphi_2}{\varphi_3}$. It follows that $\rho=\lambda w a$ with $a$ and $b$ satisfying
\begin{eqnarray}\label{con1}
a_x=b-\lambda wa^2,\\ \label{con2}
 b_x=(1+\lambda u^2)a+v-\lambda w a b.
\end{eqnarray}
Solving the above system by expanding $a,b$ as $a=\sum_{j\geq0}a_j\lambda^{j}, b=\sum_{j\geq0}b_j\lambda^{j}$ yields
\begin{eqnarray*}
a_{0x}=b_0, \quad \quad \quad \quad \quad \quad \quad \quad \quad  b_{0x}=a_0+v,\\
a_{1x}=b_1-wa_0^2,\quad \quad \quad \quad \quad \quad  b_{1x}=a_1+u^2a_0-wa_0b_0,\\
a_{ix}=b_i-w\sum_{k=0}^{i-1}a_ka_{i-k-1},\quad \quad  b_{ix}=a_i+u^2a_{i-1}-w\sum_{k=0}^{i-1}a_kb_{i-k-1}, (i\geq2).
\end{eqnarray*}We obtain, after some calculations, that
\begin{eqnarray*}
a_0=-p,\quad \quad \quad \quad \quad \quad \quad \quad \quad \quad \quad \quad \quad \quad  \ \  b_0=-p_x,\\
a_1=(1-\partial^2)^{-1}(u^2p+3wpp_x+w_xp^2), \quad \quad b_1=wp^2+a_{1x},\\
a_i=(1-\partial^2)^{-1}[w\sum_{k=0}^{i-1}a_kb_{i-k-1}+(w\sum_{k=0}^{i-1}a_ka_{i-k-1})_x-u^2a_{i-1}],\\
 b_i=a_{ix}+w\sum_{k=0}^{i-1}a_ka_{i-k-1}, \quad (i\geq2).
\end{eqnarray*}
Then an infinite sequences of conserved quantities are gotten. The first three are
\begin{eqnarray*}
\Gamma_1=-\int pw dx,\\
\Gamma_2=\int [u^2pr+wpp_xr-wp^2r_x] dx,\\
\Gamma_3=\int [a_1(3wpr_x+w_xpr-u^2r)-w^2p^3r] dx.
\end{eqnarray*}
Furthermore, we can also expanding $a,b$ as
\begin{equation*}
a=\sum_{j\geq1}a_j\lambda^{-\frac{1}{2}j}, \quad \quad \quad b=\lambda^{\frac{1}{2}}\sum_{j\geq1}b_j\lambda^{-\frac{1}{2}j},
\end{equation*}which are different from the expansions in \cite{Xue}. Taking the similar procedure as the previous, we have
\begin{eqnarray*}
a_1=uw^{-1},\quad \quad \quad \quad \quad \quad \quad \quad \quad \quad \quad \ \ b_1=u^2w^{-1}, \\
a_2=\frac{1}{2}u^{-2}v-\frac{3}{2}(uw)^{-1}u_x+w^{-2}w_x,\quad b_2=u^{-1}v-u^{-1}(u^2w^{-1})_x,\\
b_{i+1}=-u^{-1}(b_{ix}-a_{i-1}+w\sum_{k=2}^{i}a_kb_{i+2-k}),\\
a_{i+1}=\frac{1}{2}u^{-1}(b_{i+1}-a_{ix}-w\sum_{k=2}^{i}a_ka_{i+2-k}).
\end{eqnarray*}Then the first four conserved quantities may be obtained, which are
\begin{eqnarray*}
\Upsilon_1=\int udx,\\
\Upsilon_2=\frac{1}{2}\int u^{-2}vw dx,\\
\Upsilon_3=\frac{1}{4}\int u^{-5}(\frac{1}{2}u^2u_x^2+u^2wv_x+2u^4-\frac{3}{2}v^2w^2-u^2vw_x)dx,\\
\Upsilon_4=\frac{1}{2}\int [-u^{-4}(vw+v_xw_x)+2u^{-6}(uu_x(vw)_x-w^2vv_x-2vwu_x^2)\\
\hspace{1.8cm}+u^{-8}(3w^2v^2uu_x+w^3v^3)]dx.\\
\end{eqnarray*}

\subsection{Hamiltonian structure}
In this part, we will study the 3-Novikov hierarchy in the view of bi-Hamiltonian structure. Notice that the 3-Novikov  system (\ref{3ch}) is generated by the two conserved quantities $\Gamma_1, \Gamma_2$, we have the following result.
\begin{thm}The 3-Novikov equation (\ref{3ch}) is a bi-Hamiltonian system, namely, it may be written as
\begin{equation}
\left(
  \begin{array}{c}
    u \\
    v \\
    w \\
  \end{array}
\right)_t={\cal J}\left(
  \begin{array}{c}
    \frac{\delta H_{2}}{\delta u} \\
    \frac{\delta H_{2}}{\delta v} \\
    \frac{\delta H_{2}}{\delta w} \\
  \end{array}
\right)={\cal K}\left(
  \begin{array}{c}
    \frac{\delta H_{1}}{\delta u} \\
    \frac{\delta H_{1}}{\delta v} \\
    \frac{\delta H_{1}}{\delta w} \\
  \end{array}
\right),
\end{equation}where
\begin{eqnarray*}
&&{\cal J}=\left(
           \begin{array}{ccc}
             \frac{1}{2}\partial & 0 & 0 \\
             0 & 0 & 1-\partial^2 \\
             0 & \partial^2-1 & 0 \\
           \end{array}
         \right),\\
&&{\cal K}=\left(
                         \begin{array}{ccc}
                           0 & 0 & 0 \\
                           0 & \frac{3}{2}v\partial^{-1}v & -u^2-\frac{3}{2}v\partial^{-1}w \\
                           0 & u^2-\frac{3}{2}w\partial^{-1}v & \frac{3}{2}w\partial^{-1}w \\
                         \end{array}
                       \right)-2\Omega(\partial^3-4\partial)^{-1}\Omega^{*},
\end{eqnarray*}
herein
\begin{eqnarray*}
&&\Omega=(\partial u,\frac{1}{2}v\partial+\partial v,\frac{1}{2}w\partial+\partial w)^{T},\\
&&H_{1}=-\Gamma_1,\quad \quad \quad \quad \quad H_{2}=-\Gamma_2.
\end{eqnarray*}
\end{thm}

Since ${\cal J}, {\cal K}$ forms a Hamiltonian pair \cite{Xue}, one can prove the theorem easily. Hence a recursion operator for 3-Novikov hierarchy is gotten as ${\cal R}={\cal K}{\cal J}^{-1}$, and we can derive a new 3-Novikov hierarchy by taking the trivial flow as $(u,v,w)^{T}_t={\cal R}(0,v,-w)^{T}$. Then the positive flows in the hierarchy may be obtained as
\begin{equation}
\left(
  \begin{array}{c}
    u \\
    v \\
    w \\
  \end{array}
\right)
_{t_{n}}={\cal J}\left(
  \begin{array}{c}
    \frac{\delta H_{n+1}}{\delta u} \\
    \frac{\delta H_{n+1}}{\delta v} \\
    \frac{\delta H_{n+1}}{\delta w} \\
  \end{array}\right)={\cal K}\left(
  \begin{array}{c}
    \frac{\delta H_{n}}{\delta u} \\
    \frac{\delta H_{n}}{\delta v} \\
    \frac{\delta H_{n}}{\delta w} \\
  \end{array}\right),\quad n=1,2,...,
\end{equation}and infinitely many negative flows read as
\begin{equation}\label{Hn}
\left(
  \begin{array}{c}
    u \\
    v \\
    w \\
  \end{array}
\right)
_{t_{-n}}={\cal K}\left(
  \begin{array}{c}
    \frac{\delta H_{-(n+1)}}{\delta u} \\
    \frac{\delta H_{-(n+1)}}{\delta v} \\
    \frac{\delta H_{-(n+1)}}{\delta w} \\
  \end{array}\right)={\cal J}\left(
  \begin{array}{c}
    \frac{\delta H_{-n}}{\delta u} \\
    \frac{\delta H_{-n}}{\delta v} \\
    \frac{\delta H_{-n}}{\delta w} \\
  \end{array}\right),\quad n=1,2,...
\end{equation}
with the first two Hamiltonian functionals giving by $H_{-1}=-2\Upsilon_2, H_{-2}=-2\Upsilon_4$. In particular, the first negative flow in the hierarchy is obtained by using the Hamiltonian functionals $H_{-1},H_{-2}$, that is
\begin{eqnarray}
u_t-(\frac{vw}{u^3})_x=0,\nonumber\\
v_t-(\frac{v}{u^2})_{xx}+\frac{v}{u^2}=0,\\
w_t-\frac{w}{u^2}+(\frac{w}{u^2})_{xx}=0.\nonumber
\end{eqnarray}
It is worth to note that $\Upsilon_1$ and $\Upsilon_3$ are the Casimir functionals of the Hamiltonian operators  ${\cal J}$ and ${\cal K}$ respectively.

Since the structure of Hamiltonian functionals $H_ns$ in the 3-Novikov hierarchy is largely unknown, like the cases in \cite{Lene,Kang}, we will consider the homogeneous and local properties of them. Introducing $\theta=(u,v,w)^{T}$ and $X_n[\theta]=\frac{\delta H_n}{\delta \theta}$, then recursive relation in the positive direction
\begin{equation*}
  {\cal J}\frac{\delta H_{n+1}}{\delta \theta}={\cal K}\frac{\delta H_n}{\delta \theta},\quad \quad \quad  n=1,2,...,
\end{equation*}yields an infinite sequence of variational derivatives for the Hamiltonian functionals $H_n$s
\begin{equation}\label{hn1}
X_{n+1}[\theta]={\cal J}^{-1}{\cal K}X_n[\theta], \quad \quad n=1,2,....
\end{equation}Similarly, the variational derivatives for the Hamiltonian functionals $H_{-n}$s in the negative direction are given by
\begin{equation*}
X_{-(n+1)}[\theta]={\cal K}^{-1}{\cal J}X_{-n}[\theta], \quad \quad n=1,2,....
\end{equation*}
\begin{prop}
The variational derivatives $X_n[\theta]$ are homogeneous in the sense that
\begin{equation}\label{poha}
X_n[\epsilon \theta]=\epsilon^{2n-1}X_n[\theta],\quad n\geq1,
\end{equation}and
\begin{equation}
H_n[\varepsilon\theta]=\frac{1}{2n}\int X_n[\theta]\cdot\theta dx, \quad  n\geq1.
\end{equation}
\end{prop}
{\bf Proof:} When $n=1$, the formulate (\ref{poha}) holds clearly.
Now suppose (\ref{poha}) also holds for $n=k$, that is
\begin{equation*}
X_{k}[\epsilon \theta]=\epsilon^{2k-1}X_k[\theta].
\end{equation*} Then for $n=k+1$, we have
\begin{eqnarray*}
X_{k+1}[\epsilon \theta]={\cal J}^{-1}[\epsilon\theta]{\cal K}[\epsilon \theta]X_{k}[\epsilon \theta]=\epsilon^2{\cal J}^{-1}[\theta]{\cal K}[ \theta]X_k[\epsilon\theta],
\end{eqnarray*}which implies that
\begin{eqnarray*}
X_{k+1}[\epsilon \theta]=\epsilon^{2k+1}[\theta]X_{k+1}[\theta].
\end{eqnarray*}
In addition, for any $n\geq 1$, we have
\begin{equation*}
H_n[\theta]=\int^1_0\int X_n[\varepsilon\theta]\cdot\theta dxd\varepsilon=\frac{1}{2n}\int X_n[\theta]\cdot\theta dx,
\end{equation*}
then the Hamiltonian functionals $H_n$s are also homogeneous with
\begin{equation*}
H_n[\varepsilon\theta]=\varepsilon^{2n}H_n[\theta],\quad n=1,2,....
\end{equation*}

The recursive formula for $H_n$s yields infinitely many Hamiltonian functionals in the positive direction, and $H_1$ and $H_2$ are local. However, $H_n, n\geq3$ becomes nonlocal. For example $H_3=-\Gamma_3$, which is shown to be nonlocal.

\begin{prop}The variational derivatives $X_{-n}[\theta]{\rm s}$ satisfy
\begin{equation}\label{neha}
X_{-n}[\epsilon \theta]=\epsilon^{1-2n}X_{-n}[\theta],\quad n=1,2,....
\end{equation}while
\begin{equation}\label{hom2}
H_{-n}[\theta]=\frac{1}{2-2n}\int X_{-n}[\theta]\cdot\theta dx,
\end{equation}and $H_{-n}{\rm s}$ are all local.
\end{prop}The formulae (\ref{neha}) and (\ref{hom2}) may be proven by taking the process before, and we will prove the local property of $H_{-n}$s below.

\begin{lem}(\cite{Lene,Kang,Pjo})
If a differential function $M[\theta]$ satisfies
\begin{equation*}
\int M[\theta] dx=0
\end{equation*}for all $\theta$, then there exists a unique differential function $N[\theta]$ up to addition of a constant such that $M[\theta]$ is the total x-derivative $M[\theta]=(N[\theta])_x$.
\end{lem}
Introducing
\begin{eqnarray*}
&&X_{-k}[\theta]=(A_k,B_k,C_k)^{T} \\
&&E_k=(\partial^3-4\partial)^{-1}(u\partial,\frac{3}{2}v\partial+\frac{1}{2}v_x,\frac{3}{2}w\partial+\frac{1}{2}w_x) X_{-k}[\theta], \quad  k\geq1.
\end{eqnarray*}
When $n=1$, $X_{-1}[\theta]$ is local since
\begin{equation*}
X_{-1}[\theta]=(2\frac{vw}{u^3},-\frac{w}{u^2},-\frac{v}{u^2})^{T}.
\end{equation*}Now suppose $X_{-k}[\theta]$ is local for $n=k$. Then for $n=k+1$, we have
\begin{equation*}
X_{-(k+1)}[\theta]={\cal K}^{-1}{\cal J}X_{-k}[\theta]=({\cal K}^{-1}{\cal J})^{k}X_{-1}[\theta],
\end{equation*}which is equal to
\begin{equation}
{\cal K}X_{-(k+1)}[\theta]={\cal J}X_{-k}[\theta].
\end{equation}This shows that
\begin{eqnarray}\label{local1}
&&E_{k+1}=\frac{1}{4u}A_{k}, \\ \label{local2}
&&\frac{3}{2}v\partial^{-1}(vB_{k+1}-wC_{k+1})-u^2C_{k+1}+(3v\partial+2v_x)E_{k+1}=(1-\partial^2)C_{k}, \\ \label{local3}
&&u^2B_{k+1}-\frac{3}{2}w\partial^{-1}(vB_{k+1}-wC_{k+1})+(3w\partial+2w_x)E_{k+1}=(\partial^2-1)B_{k}.
\end{eqnarray}

Then we will prove the local property of $X_{-(k+1)}$ in two steps. The first step is to prove that $B_{k+1}$ and $C_{k+1}$ are local. Since $A_k,B_k,C_k$ are all local,
we can obtain immediately from (\ref{local2}) and (\ref{local3}) that $B_{k+1}$ and $C_{k+1}$ are local, if there exist a differential function $M_k$ such that
\begin{eqnarray*}
&&vB_{k+1}-wC_{k+1}=\frac{w}{u^2}(1-\partial^2)C_k+\frac{v}{u^2}(\partial^2-1)B_k-\frac{3vw}{2u^2}\partial\frac{A_k}{u}-\frac{(vw)_x}{2u^3}A_k\\
&&\hspace{2.8cm}=M_{kx}.
\end{eqnarray*}Then according to the Lemma 1, we only need to prove
\begin{equation*}
Y_1=\int [\frac{w}{u^2}(1-\partial^2)C_k+\frac{v}{u^2}(\partial^2-1)B_k-\frac{3vw}{2u^2}\partial\frac{A_k}{u}-\frac{(vw)_x}{2u^3}A_k]dx=0.
\end{equation*}
In fact
\begin{eqnarray*}
&&Y_1=\int [\frac{w}{u^2}(1-\partial^2)C_k+\frac{v}{u^2}(\partial^2-1)B_k-\frac{3vw}{2u^2}\partial\frac{A_k}{u}-\frac{(vw)_x}{2u^3}A_k]dx\\
&&\hspace{0.4cm}=\int [C_k(1-\partial^2)\frac{w}{u^2}+B_k(\partial^2-1)\frac{v}{u^2}+A_k(\frac{vw}{u^3})_x]dx\\
&&\hspace{0.4cm}=\int\left(
                   \begin{array}{c}
                     A_k \\
                     B_k \\
                     C_k \\
                   \end{array}
                 \right)\cdot {\cal J}\left(
                                        \begin{array}{c}
                                          2\frac{vw}{u^2} \\
                                          -\frac{W}{u^2} \\
                                          -\frac{v}{u^2} \\
                                        \end{array}
                                      \right)dx\\
&&\hspace{0.4cm}=\int X_{-k}[\theta]\cdot {\cal J}X_{-1}[\theta]dx.
\end{eqnarray*}On the other hand, using the recursion relation, we have
\begin{eqnarray*}
&&Y_1=\int({\cal K}^{-1}{\cal J})^{k-1}X_{-1}[\theta]\cdot {\cal J}X_{-1}[\theta]dx\\
&&\hspace{0.4cm}=-\int X_{-1}[\theta]\cdot{\cal J}({\cal K}^{-1}{\cal J})^{k-1}X_{-1}[\theta]dx\\
&&\hspace{0.4cm}=-\int X_{-1}[\theta]\cdot({\cal J}{\cal K}^{-1})^{k-1}{\cal J}X_{-1}[\theta]dx\\
&&\hspace{0.4cm}=-\int ({\cal K}^{-1}{\cal J})^{k-1}X_{-1}[\theta]\cdot{\cal J}X_{-1}[\theta]dx\\
&&\hspace{0.4cm}=-\int X_{-k}[\theta]\cdot {\cal J}X_{-1}[\theta]dx.
\end{eqnarray*}Therefore $Y_1=0$, and hence $B_{k+1}$ and $C_{k+1}$ are local.

The next step is to prove that $A_{k+1}$ is local. From (\ref{local1}), we infer that
\begin{equation*}
A_{k+1x}=\frac{1}{u}[(\partial^3-4\partial)\frac{A_k}{4u}-(3v\partial+2v_x)B_{k+1}-(3w\partial+2w_x)C_{k+1}].
\end{equation*}Notice that $B_{k+1}$ and $C_{k+1}$ are all local, so $A_{k+1}$ is local if the right part of the above equality is a total $x$-derivative $N_{kx}$ for a differential function $N_k$. That is to say,  $A_{k+1}$ is local if
\begin{equation*}
Y_2=\int (\frac{1}{u}[(\partial^3-4\partial)\frac{A_k}{4u}-(3v\partial+2v_x)B_{k+1}-(3w\partial+2w_x)C_{k+1}])dx=0.
\end{equation*}

\begin{lem}Define
\begin{equation*}
{\cal D}=\left(
  \begin{array}{cc}
    \frac{3}{2}v\partial^{-1}v & -u^2-\frac{3}{2}v\partial^{-1}w \\
    u^2-\frac{3}{2}w\partial^{-1}v & \frac{3}{2}w\partial^{-1}w \\
  \end{array}
\right),
\end{equation*}we have
\begin{equation*}
{\cal D}^{-1}=\frac{1}{u^2}\left(
                            \begin{array}{cc}
                              \frac{3}{2}w\partial^{-1}w & u^2+\frac{3}{2}w\partial^{-1}v \\
                              \frac{3}{2}v\partial^{-1}w-u^2 & \frac{3}{2}v\partial^{-1}v \\
                            \end{array}
                          \right)
\frac{1}{u^2}.
\end{equation*}
\end{lem}
To make the expressions compact, we introduce some new notations as:
\begin{eqnarray*}
Z_1=(1-\partial^2)C_k-(3v\partial+2v_x)\frac{A_k}{4u},Z_2=(\partial^2-1)B_k-(3vw\partial+2w_x)\frac{A_k}{4u},\\
Z_3=\frac{v_x}{u^3}-\frac{3}{2}\frac{u_xv}{u^4}-\frac{3}{2}\frac{wv^2}{u^4},\quad \quad \quad \quad Z_4=-\frac{w_x}{u^3}+\frac{3}{2}\frac{u_xw}{u^4}-\frac{3}{2}\frac{vw^2}{u^5}.
\end{eqnarray*}
Using the Lemma 2 to solve $B_{k+1}$ and $C_{k+1}$ from (\ref{local2}) and (\ref{local3}), we arrive at
\begin{eqnarray*}
&&Y_2=\int (\frac{1}{u}[(\partial^3-4\partial)\frac{A_k}{4u}-(3v\partial+2v_x)B_{k+1}-(3w\partial+2w_x)C_{k+1}])dx\\
&&\hspace{0.4cm}=\int [-\frac{A_k}{4u}(\partial^3-4\partial)\frac{1}{u}+B_{k+1}(\frac{v_x}{u}-\frac{3u_xv}{u^2})+C_{k+1}(\frac{w_x}{u}-\frac{3u_xw}{u^2})]dx\\
&&\hspace{0.4cm}=\int   [-\frac{A_k}{4u}(\partial^3-4\partial)\frac{1}{u}+(\frac{v_x}{u^3}-\frac{3u_xv}{2u^4})[\frac{3}{2}w\partial^{-1}(\frac{wZ_1+vZ_2}{u^2})+Z_2]\\
&&\hspace{0.8cm}+(\frac{w_x}{u^3}-\frac{3u_xw}{2u^4})[\frac{3}{2}v\partial^{-1}(\frac{wZ_1+vZ_2}{u^2})-Z_1]]dx\\
&&\hspace{0.4cm}=\int[-\frac{A_k}{4u}(\partial^3-4\partial)\frac{1}{u}+Z_2Z_3+Z_1Z_4]dx\\
&&\hspace{0.4cm}=\int (\frac{A_k}{4u}[-(\partial^3-4\partial)\frac{1}{u}+(3w\partial+2w_x)Z_3+(3v\partial+2v_x)Z_4]\\
&&\hspace{0.8cm}+B_k(\partial^2-1)Z_1+C_k(1-\partial^2)Z_2)dx\\
&&\hspace{0.4cm}=\int\left(
                   \begin{array}{c}
                     A_k \\
                     B_k \\
                     C_k \\
                   \end{array}
                 \right)\cdot {\cal J}\left(
                                        \begin{array}{c}
                                          \frac{3wv_x-3vw_x}{2u^4}-\frac{15v^2w^2}{4u^6}+\frac{u_{xx}}{2u^3}-\frac{3u_x^2}{4u^4}+\frac{1}{u^2} \\
                                          \frac{w_x}{u^3}-\frac{3wu_x}{2u^4}+\frac{3vw^2}{2u^5} \\
                                          -\frac{v_x}{u^3}+\frac{3vu_x}{2u^4}+\frac{3v^2w}{2u^5} \\
                                        \end{array}
                                      \right)dx\\
&&\hspace{0.4cm}=\int[-2X_{-k}[\theta]\cdot {\cal J}\frac{\delta\Upsilon_3}{\delta\theta}]dx\\
&&\hspace{0.4cm}=\int2{\cal J}({\cal K}^{-1}{\cal J})^{k-1}X_{-1}[\theta]\cdot \frac{\delta\Upsilon_3}{\delta\theta}dx\\
&&\hspace{0.4cm}=\int-2({\cal K}^{-1}{\cal J})^{k}X_{-1}[\theta]\cdot {\cal K}\frac{\delta\Upsilon_3}{\delta\theta}dx\\
&&\hspace{0.4cm}=\int-2({\cal K}^{-1}{\cal J})^{k}X_{-1}[\theta]\cdot (0,0,0)^{T}dx\\
&&\hspace{0.4cm}=0.
\end{eqnarray*}Therefore $A_{k+1}$ is local. Consequently, we prove $X_{-n}{\rm s}$ are all local. Then using the Lemma 4.4 in \cite{Kang} (see also \cite{Lene,Pjo}), $H_{-n}{\rm s}$ are found to be local.

\section{Relationship with a 3-component CH type system}
As pointed out in \cite{Popowicz},  the 3-CH type system (\ref{3eqq}) is reciprocal linked to the first negative flow in a mYJ hierarchy. Since
the spectral problem (\ref{thr2}) is gauge linked to the spectral problem (\ref{lax1}), it would seem to be a reasonable guess that the 3-CH type system (\ref{3eqq}) is equal to the 3-Novikov system (\ref{3ch}).

In fact the 3-CH type system (\ref{3eqq}) may be rewritten as
\begin{eqnarray}\label{po}
\begin{array}{l}m_{1t}+u_2gm_{1x}-m_{3}(u_{2x}f-u_2g)-m_1(3u_2f-m_3u_2)=0,\\
m_{2t}+u_2gm_{2x}+m_{2}(3u_{2x}g+m_3u_2)=0, \\
m_{3t}+u_2gm_{3x}-m_3(2u_2f+u_{2x}g-m_3u_2)=0, \\
m_{2}=u_{2}-u_{2xx}, \quad m_1-m_{3x}=g-g_{xx}, \quad f=m_3-g_x, \end{array}
\end{eqnarray} then a directly calculation shows that (\ref{po}) is connected to (\ref{3ch}) via
\begin{equation}
u=(m_2m_3)^{\frac{1}{2}},\quad v=m_2,\quad w=m_1-m_{3x},\quad p=u_2,\quad r=g.
\end{equation}

\section{A reciprocal transformation for the 3-Novikov system}
\subsection{A reciprocal transformation}
Although many CH type systems are completely integrable, they have some nonstandard features such as the DT, the B\"{a}cklund transformation  and the weak Painlev\'{e} property \cite{DHH,Gilson}.
To study the Painlev\'e behaviour of the 3-Novikov system (\ref{3ch}), we can relate it with a equation displaying the standard (strong) Painlev\'e test of WTC \cite{hone5}, which is easy to construct the DT and the B\"{a}cklund transformation. Our strategy is to use the steps in \cite{Li2}, and we will connect the 3-Novikov system  (\ref{3ch}) with a negative flow in a mYJ hierarchy.

The 3-Novikov system  (\ref{3ch}) has a conserved density $u$, and the correspondence conservation law reads
\begin{equation*}
u_t=(-upr)_x,
\end{equation*}which allows a reciprocal transformation
\begin{equation}\label{rt}
dy=udx-upr dt,\quad \quad \quad \quad d\tau=dt.
\end{equation}Set $\mu=\lambda^{\frac{1}{2}}$ and define $\psi_2=\mu\frac{u^2}{v}\varphi_1+\frac{1}{\mu}\varphi_3$. Then, under change of variables, we may rewrite the spectral problem (\ref{lax1}) as
\begin{eqnarray}\label{med1}
 \begin{array}{l}\varphi_{1yy}+\frac{u_y}{u}\varphi_{1y}-\frac{1}{u^2}\varphi_1-\mu \frac{v}{u^2}\psi_2=0, \\
  \psi_{2y}-\mu \frac{u^2}{v}\varphi_{1y}-\mu[(\frac{u^2}{v})_y+\frac{w}{u}]\varphi_1=0.\end{array}
\end{eqnarray} Now, introducing the gauge transformation
\begin{equation*}
\varphi_1= \frac{v}{u^2}e^{-\partial_y^{-1}(\frac{vw}{u^3})}\phi_1, \quad \quad \quad \psi_2= e^{-\partial_y^{-1}(\frac{vw}{u^3})}\phi_2,
\end{equation*}
the spectral problem (\ref{med1}) may be converted to
\begin{eqnarray}\label{sca1}
\begin{array}{l}\phi_{1yy}-Q_2\phi_{1y}-Q_1\phi_1=\mu \phi_2,\\
\phi_{2y}-Q_3\phi_2=\mu \phi_{1y},\end{array}
\end{eqnarray}where
\begin{eqnarray*}
&&Q_1=(3\frac{v_y}{v}-6\frac{u_y}{u}+\frac{w_y}{w}-\frac{vw}{u^{3}})\frac{vw}{u^{3}}+3\frac{u_yv_y}{uv}+\frac{1+2uu_{yy}-4u_y^2}{u^2}
-\frac{v_{yy}}{v}, \\
&&Q_2=2\frac{vw}{u^{3}}-2\frac{v_y}{v}+3\frac{u_y}{u}, \\
&&Q_3=\frac{vw}{u^{3}}.
\end{eqnarray*}
It is easy to check that the auxiliary problem in (\ref{Laxpair}) is transformed to
\begin{eqnarray}\label{sca3}
 \begin{array}{l}\phi_{1\tau}=\frac{1}{\mu}q_1\phi_2+\frac{1}{3\mu^2}\phi_1, \\
  \phi_{2\tau}=\frac{1}{\mu}(q_2\phi_{1y}+[1-q_{2y}+(Q_3-Q_2)q_2]\phi_1)+(q_1-\frac{2}{3\mu^2})\phi_2,\end{array}
\end{eqnarray}where
\begin{equation}
q_1=p\frac{u^2}{v},\quad \quad \quad \quad \quad \quad q_2=\frac{rv}{u}.
\end{equation}

For the convenience of constructing exact solutions of the 3-Novikov equation, let us rewrite the above Lax pair in scalar form. Eliminating $\phi_2$ from the systems (\ref{sca1}) and (\ref{sca3}), we obtain
\begin{eqnarray}\label{assnovikov1}
 &&\phi_{1yyy}+u_1\phi_{1yy}+(v_1+u_{1y})\phi_{1}+(w_1+v_{1y})\phi_1=\lambda\phi_{1y}, \\ \label{assnovikov2}
&&\phi_{1\tau}-\frac{1}{\lambda}(q_1\phi_{1yy}+(u_1+Q_3)q_1\phi_{1y}-\chi\phi_1)=0,
\end{eqnarray}where

\begin{equation}\label{miu}
\left(
  \begin{array}{c}
    u_1 \\
    v_1 \\
    w_1 \\
  \end{array}
\right)=\left(
          \begin{array}{c}
            -Q_2-Q_3 \\
            Q_2Q_3-Q_1+Q_{3y} \\
            Q_1Q_3-(Q_2Q_3)_y-Q_{3yy} \\
          \end{array}
        \right)
\end{equation}
with
\begin{equation*}
\chi=q_{1yy}+(u_1+3Q_3)q_{1y}+[(u_1+2Q_3)Q_3+Q_{3y}]q_1+\frac{2}{3}.
\end{equation*}
Then the compatibility condition for the Lax representation (\ref{assnovikov1}-\ref{assnovikov2}) yields the associated 3-Novikov equation
\begin{eqnarray}\label{associated3n}
\begin{array}{l}
\left(
  \begin{array}{c}
    u_1 \\
    v_1\\
    w_1\\
  \end{array}
\right)_{\tau}=\left(
          \begin{array}{c}
           -2q_{1y}, \\
           -q_{1yy}-u_1 q_{1y}-2(Q_3q_1)_y,\\
            -[Q_3q_{1y}+q_1(u_1 Q_3+2Q_3^2-Q_{3y})]_y,\\
          \end{array}
        \right),\ \left(
                   \begin{array}{c}
                     s_1 \\
                     s_2 \\
                   \end{array}
                 \right)=0,\end{array}
\end{eqnarray}
where
\begin{eqnarray*}
&&s_1=q_{1yy}+q_1(2u_1 Q_3+3Q_3^2+v_1)+q_{1y}(3Q_3+u_1)+1,\\
&&s_2=w_1+Q_3(v_1+Q_3^2-3Q_{3y}-u_{1y})+Q_{3yy}+(Q_3^2-Q_{3y})u_1.
\end{eqnarray*}

Furthermore, one can also gain the associated 3-Novikov system (\ref{associated3n}) by applying the reciprocal transformation (\ref{rt}) to the 3-Novikov system (\ref{3ch}) directly.  Now, we claim that the 3-Novikov equation (\ref{3ch}) and the Lax pair (\ref{Laxpair}) is reciprocal transformed to the associated equation (\ref{associated3n}) and  the Lax pair (\ref{assnovikov1}-\ref{assnovikov2}) respectively. More precisely, we have:
\begin{prop}
The 3-Novikov equation (\ref{associated3n}) may be changed to the associated equation (\ref{associated3n}) by the Liouville transformation

\begin{eqnarray}\label{Liouville}
\left\{\begin{array}{rl}
&y=I(x,\theta^{(n)})=\int_{-\infty}^{x} u(\nu)d\nu,\\
&\left(
   \begin{array}{c}
     u_1(y) \\
     v_1(y) \\
     w_1(y) \\
   \end{array}
 \right)=\left(
           \begin{array}{c}
             P_1(x,\theta^{(n)}), \\
             P_2(x,\theta^{(n)}), \\
             P_3(x,\theta^{(n)}), \\
           \end{array}
         \right),\end{array} \right.
\end{eqnarray}where
\begin{eqnarray*}
P_1=2\frac{v_x}{vu}-3\frac{u_x}{u^2}-3\frac{vw}{u^3},\\
P_2=\frac{v_{xx}-v}{u^2v}-4\frac{u_xv_x}{u^3v}-\frac{4wv_x+2uu_{xx}-6u_x^2}{u^4}+6\frac{vwu_x}{u^5}+3\frac{v^2w^2}{u^6},\\
P_3=\frac{v(w-w_{xx})}{u^5}+\frac{4vw_xu_x+2vwu_{xx}}{u^6}-\frac{6vwu_x^2+3v^2ww_x+vw^2v_x}{u^7}\\
\hspace{1.0cm}+\frac{6v^2w^2u_x}{u^8}-\frac{v^3w^3}{u^9},
\end{eqnarray*}
with $v=p-p_{xx}, w=r-r_{xx}$.
\end{prop}

It is worth to note that the associated 3-Novikov system  passes the Painlev\'e test. Powers of the leader terms for $u_1,v_1,w_1,q_1,Q_3$ are $-1,-2,-2,-1,-1$ respectively, and the resonances are $j=-2,-1,1,2,3,4,5$.

The spectral problem (\ref{assnovikov1}) may be rewritten as the Lax operator for a mYJ hierarchy
\begin{equation}
L\phi_1=\lambda \phi_1,\quad \quad \quad L=\partial_y^2+u_1\partial_y+v_1+\partial_y^{-1}w_1,
\end{equation}which is just a member in the constrained modified KP hierarchy \cite{Oevel,Kono}. It can reduce to that of the mKdV equation and the KdV hierarchy as $Q_1=Q_3=0$ and $Q_2=Q_3=0$ respectively. We claim that the associated 3-Novikov equation is a reduction of  the first negative flow in the mYJ hierarchy.

Notice that the mYJ hierarchy admits a Hamiltonian pair
\begin{eqnarray*}
 {\cal J}_1 &=& \left(
                   \begin{array}{ccc}
                     0 & 0 & 2\partial_y \\
                     0 & 2\partial_y & \partial_y^2+u_1\partial_y \\
                     2\partial_y & -\partial_y^2+\partial_y u_1 & 0 \\
                   \end{array}
                 \right),\\
  {\cal K}_1 &=& \left(
                   \begin{array}{ccc}
                     6\partial_y & * & * \\
                     4u_1\partial_y & 2\partial_y^3+2u_1\partial_y u_1+\partial_y v_1+v_1\partial_y & * \\
                    2\partial_y^3-2\partial_y u_1\partial_y+2v_1\partial_y & \chi_1 & \chi_2 \\
                   \end{array}
                 \right),
\end{eqnarray*}where
\begin{eqnarray*}
&&\chi_1=2w_1\partial_y+\partial_y w_1-(\partial_y^3-\partial_y u_1\partial_y+v_1\partial_y)(\partial_y-u_1),\\
&&\chi_2=\partial_y u_1w_1+u_1w_1\partial_y+w_1\partial_y^2-\partial_y^2 w_1
\end{eqnarray*}and the omitted terms are determined by skew-symmetry.
Then a recursion operator for the mYJ  hierarchy is obtained as ${\cal R}={\cal K}_1{\cal J}_1^{-1}$, and the first negative flow in the correspondence hierarchy are obtained as
\begin{equation}\label{firstflow}
\left(
  \begin{array}{c}
    u_1 \\
    v_1 \\
    w_1 \\
  \end{array}
\right)_\tau={\cal J}_1\left(
                              \begin{array}{c}
                               A \\
                               B \\
                               C\\
                              \end{array}
                            \right), \quad \quad \quad {\cal K}_1\left(
                              \begin{array}{c}
                               A \\
                               B \\
                               C\\
                              \end{array}
                            \right)=0,
\end{equation}where $A=A(y,\tau),B=B(y,\tau),C=C(y,\tau)$. To find the relation between the associated 3-Novikov system (\ref{associated3n}) and the negative flow (\ref{firstflow}), we can take
\begin{equation}\label{assume}
A=-Q_3q_{1y}-Q_3^2q_1,\quad \quad \quad
B=-Q_3q_1,\quad \quad \quad
 C=-q_1.
\end{equation}
Then the negative flow (\ref{firstflow}) is changed to
\begin{eqnarray}
\left(
  \begin{array}{c}
    u_1 \\
    v_1 \\
    w_1\\
  \end{array}
\right)_{\tau}=\left(
          \begin{array}{c}
           -2q_{1y}, \\
           -q_{1yy}-u_1 q_{1y}-2(Q_3q_1)_y,\\
            -[Q_3q_{1y}+q_1(u_1 Q_3+2Q_3^2-Q_{3y})]_y,\\
          \end{array}
        \right),\quad \left(
                         \begin{array}{c}
                          z_1 \\
                           z_2 \\
                           z_3 \\
                         \end{array}
                       \right)=0,
\end{eqnarray}
where
\begin{eqnarray*}
z_1=-2s_1,\\
z_2=-s_{1yy}+(Q_3-u_1)s_{1y}-2p_1s_{2y}-3p_{1y}s_2,\\
z_3=(Q_3^2-2Q_{3y}+Q_3u_1)s_{1y}-Q_3s_{1yy}+p_1s_{2yy}+(2p_{1y}-Q_3p_1-u_1 p_1)s_{2y}\\
\hspace{1.0cm}-(p_1u_{1y}+2p_{1y}u_1+3(p_1Q_3)_y)s_2.
\end{eqnarray*}Thus the associated 3-Novikov system (\ref{associated3n}) is a reduction of the first negative flow (\ref{firstflow}) in the mYJ hierarchy, since $s_1=0,s_2=0$ yields $z_1=z_2=z_3=0$.

\subsection{Hamiltonian structure behavior under the Liouville transformation}
According to \cite{Brunelli}, if two soliton equations are linked by a Liouville transformation, Hamiltonian structures and conserved quantities of them can be related. In this part we will consider the Hamiltonian structures of the 3-Novikov system (\ref{3ch}) under the Liouville transformation (\ref{Liouville}). To this end, let $\vartheta=(u_1,v_1,w_1)^{T}$.  Then from the point of view of Hamiltonian structures, we have
\begin{eqnarray}\label{he1}
\theta_t={\cal B}(\theta)\frac{\delta H}{\delta \theta}={\cal B}(\theta)E_{\theta}h,\\ \label{he2}
\vartheta_t=\tilde{{\cal B}}(\vartheta)\frac{\delta \tilde{H}}{\delta \vartheta}=\tilde{{\cal B}}(\vartheta)E_{\vartheta}\tilde{h},
\end{eqnarray}where
\begin{equation*}
H=\int h(x,\theta^{(n)})dx,\quad \tilde{H}=\int \tilde{h}(y,\vartheta^{(n)})dy.
\end{equation*}Herein $E_{\theta},E_{\vartheta}$ are the corresponding Euler operators, and $H[\theta^{(n)}]= \tilde{H}[\vartheta^{(n)}]$. Defining  $\Lambda(\vartheta,\theta)=\vartheta-(P_1,P_2,P_3)^{T}$, hence it is easy to see that
\begin{equation}\label{fieldre}
\vartheta_t=-T_1\theta_t, \quad  T_1=\Lambda_{\theta},
\end{equation}where $\Lambda_{\theta}$ is Frech\'{e}t derivative for the vector variable. Then a direct computation shows that
\begin{equation*}
T_1=\left(
                                   \begin{array}{ccc}
                                     u_{1y}I'[u]-P'_1[u] & u_{1y}I'[v]-P'_1[v] & u_{1y}I'[w]-P'_1[w]\\
                                     v_{1y}I'[u]-P'_2[u] & v_{1y}I'[v]-P'_2[v] & v_{1y}I'[w]-P'_2[w] \\
                                     w_{1y}I'[u]-P'_3[u] & w_{1y}I'[v]-P'_3[v] & w_{1y}I'[w]-P'_3[w] \\
                                   \end{array}
                                 \right).
\end{equation*}Furthermore, the action of Euler operator under a change of variables is given by
\begin{equation}\label{func}
E_{\theta}h=T_2E_{\vartheta}\tilde{h},
\end{equation}where
\begin{equation*}
T_2=\left(
      \begin{array}{ccc}
        P^{'\dag}_{1,u}(I_x)- I^{'\dag}_{u}(P_{1x}) &  P^{'\dag}_{2,u}(I_x)- I^{'\dag}_{u}(P_{2x}) &  P^{'\dag}_{3,u}(I_x)- I^{'\dag}_{u}(P_{3x}) \\
        P^{'\dag}_{1,v}(I_x)- I^{'\dag}_{v}(P_{1x}) &  P^{'\dag}_{2,v}(I_x)- I^{'\dag}_{v}(P_{2x}) &  P^{'\dag}_{3,v}(I_x)- I^{'\dag}_{v}(P_{3x}) \\
        P^{'\dag}_{1,w}(I_x)- I^{'\dag}_{w}(P_{1x}) &  P^{'\dag}_{2,w}(I_x)- I^{'\dag}_{w}(P_{2x}) &  P^{'\dag}_{3,w}(I_x)- I^{'\dag}_{w}(P_{3x}) \\
      \end{array}
    \right).
\end{equation*}
\begin{lem}
Under the transformation (\ref{Liouville}), we have the following formulaes:
\begin{equation*}
T_1={\cal O}{\rm diag}(u^{-1},v^{-1},vu^{-3}),\quad T_2=-{\rm diag}(1,uv^{-1},vu^{-2}){\cal O}^{\dag},
\end{equation*}where
\begin{equation*}
\hspace{-2.0cm}{\cal O}=\left(
      \begin{array}{ccc}
       u_{1y}\partial_y^{-1}+3\partial_y+u_1-6Q_3 & -2\partial_y+3Q_3 & 3\\
       v_{1y}\partial^{-1}_y+2\partial_y^2+2u_1\partial_y+2v_1-4Q_3u_1 & (Q_3-u_1)\partial_y+2Q_3u_1-\partial_y^2 & 2u_1\\
       w_{1y}\partial^{-1}_y+3w_1-2\chi_3Q_3 & \partial_y Q_3^2-w_1+Q_2Q_3^2+Q_3Q_{3y} & \chi_3\\
      \end{array}
    \right)
\end{equation*}with
\begin{equation*}
\chi_3=\partial_y^2-\partial_yu_1+v_1.
\end{equation*}
\end{lem}
\begin{lem}Under the reciprocal transformation (\ref{rt}),  the following identities hold:
\begin{equation}\label{ide1}
  \frac{1}{v}(1-\partial_x^2)\frac{v}{u^2}=\Theta_1\equiv Q_1-(\partial_y-Q_2+Q_3)(\partial_y+Q_3),
\end{equation}
and
\begin{equation}\label{ide2}
\frac{1}{u^2}(\partial_x^3-4\partial_x)\frac{1}{u}=\Theta_2\equiv(\partial_y-Q_2)\partial_y(\partial_y+Q_2)-2Q_1\partial_y-2\partial_y Q_1.
\end{equation}
\end{lem}
The two Lemmas above can be proved through a straightforward computation. Hence the main results can be summarized as:

\begin{thm}The associated 3-Novikov system is a bi-Hamiltonian system, namely, it can be written as
\begin{equation}
\left(
  \begin{array}{c}
    u_1 \\
    v_1 \\
    w_1 \\
  \end{array}
\right)_t={\cal K}_1{\cal J}_1^{-1}{\cal K}_1\left(
                            \begin{array}{c}
                              \frac{\delta \tilde{H_2}}{\delta u_1} \\
                              \frac{\delta \tilde{H_2}}{\delta v_1} \\
                              \frac{\delta \tilde{H_2}}{\delta w_1} \\
                            \end{array}
                          \right)={\cal K}_1\left(
                            \begin{array}{c}
                              \frac{\delta \tilde{H_1}}{\delta u_1} \\
                              \frac{\delta \tilde{H_1}}{\delta v_1} \\
                              \frac{\delta \tilde{H_1}}{\delta w_1} \\
                            \end{array}
                          \right),
\end{equation}
where
\begin{eqnarray*}
  \tilde{H_1} &=& \int Q_3q_1dy, \\
  \tilde{H_2} &=& \int [Q_3q_1(q_1q_{2y}-q_2q_{1y}+q_1q_2(Q_2-2Q_3))-q_1q_2] dy.
\end{eqnarray*}
\end{thm}
{\bf Proof:} Substituting (\ref{fieldre}-\ref{func}) into (\ref{he1}-\ref{he2}),  a Hamiltonian pair for the associated 3-Novikov system is obtained as
\begin{equation}\label{Ham1}
\tilde{{\cal J}}=-T_1{\cal J}T_2,\quad \tilde{{\cal K}}=-T_1{\cal K}T_2.
\end{equation} Hamiltonian functionals of the 3-Novikov system and the associated 3-Novikov system connected by the formula (\ref{func}).

To obtain bi-Hamiltonian structure of the associated 3-Novikov system, we should calculate $\tilde{{\cal J}}$ and $\tilde{{\cal K}}$ in the new variable $y$.
Using conjugation of operator to the identity (\ref{ide1}), we can easily check that
\begin{equation}\label{ide3}
\frac{v}{u^{3}}(\partial_x^2-1)\frac{u}{v}=(\partial_y-Q_3)(\partial_y+Q_2-Q_3)-Q_1.
\end{equation}Let us substitute the equalities (\ref{ide1}) and (\ref{ide3}) into the first equality in (\ref{Ham1}). Then, through tedious calculations, we get

\begin{eqnarray*}
\tilde{{\cal J}}={\cal O}{\rm diag}(u^{-1},v^{-1},vu^{-3}){\cal J}{\rm diag}(1,uv^{-1},vu^{-2}){\cal O}^{\dag}\nonumber\\
\hspace{0.5cm}={\cal O}  \left(
                             \begin{array}{ccc}
                               \frac{1}{2}\partial_y & 0 & 0 \\
                               0 & 0 & \Theta_1\\
                               0 & -\Theta_{1}^{\dag} & 0 \\
                             \end{array}
                           \right){\cal O}^{\dag}\\
\hspace{0.5cm}={\cal K}_1{\cal J}_1^{-1}{\cal K}_1.
\end{eqnarray*}On the other hand, introducing
\begin{equation*}
 \quad {\cal P}=(\partial_y,\frac{3}{2}\partial_y-\frac{1}{2}Q_2+Q_3,\frac{3}{2}Q_3\partial_y+\frac{1}{2}Q_2Q_3+Q_{3y}-Q_3^2)^{T}.
\end{equation*}Then the second equality in (\ref{Ham1}) may be changed to
\begin{eqnarray*}
\hspace{-0.4cm}\tilde{{\cal K}}
={\cal O}{\rm diag}(u^{-1},v^{-1},vu^{-3}){\cal K}{\rm diag}(1,uv^{-1},vu^{-2}){\cal O}^{\dag}\\
={\cal O}\left(
          \begin{array}{ccc}
            0 & 0 & 0 \\
            0 & \frac{3}{2}\partial_y^{-1} & -1-\frac{3}{2}\partial_y^{-1}Q_3 \\
            0 & 1-\frac{3}{2}Q_3\partial_y^{-1} &\frac{3}{2}Q_3\partial_y^{-1}Q_3  \\
          \end{array}
        \right){\cal O}^{\dag}-2{\cal O}{\cal P} \Theta_2^{-1}{\cal P}^{\dag}{\cal O}^{\dag}\nonumber\\
={\cal O}\left(
          \begin{array}{ccc}
            0 & 0 & 0 \\
            0 & \frac{3}{2}\partial_y^{-1} & -1-\frac{3}{2}\partial_y^{-1}Q_3 \\
            0 & 1-\frac{3}{2}Q_3\partial_y^{-1} & \frac{3}{2}Q_3\partial_y^{-1}Q_3  \\
          \end{array}
        \right){\cal O}^{\dag}-\left(
                                              \begin{array}{c}
                                                0 \\
                                                1 \\
                                                -Q_3 \\
                                              \end{array}
                                            \right)
\frac{\Theta_2^{\dag}}{2}\left(
                                              \begin{array}{c}
                                                0 \\
                                                1 \\
                                                -Q_3 \\
                                              \end{array}
                                            \right)^{T}
\nonumber\\
={\cal K}_1
\end{eqnarray*}by using the identity (\ref{ide2}) and ${\cal O}{\cal P}=\frac{1}{2}(0,\Theta_2,-Q_3\Theta_2)^{T}$.

Furthermore, since the Hamiltonian functionals of the two hierarchy are connected by the relation  $H[\theta^{(n)}]= \tilde{H}[\vartheta^{(n)}]$, we can easy to find the relationship between the two hierarchies.

\section{A limit system}
The limits of the CH type equations might also contain some important models. For example, the Hunter-Saxton equation, which can describe wave motion in a nematic liquid crystal \cite{Hunt}, may be consider as a limit of the CH equation \cite{Dai}. The Ostrovsky equation, which appears as the description of high-frequency waves in a relaxing medium \cite{Vak}, can be obtained as a short wave limit of the DP equation \cite{Wang}. In this section, we will consider a limit of the 3-Novikov system (\ref{3ch}).

Let us consider the transformation
\begin{equation}\label{limt}
 x\rightarrow \epsilon x, \quad   t\rightarrow \epsilon t, \quad u\rightarrow \epsilon^{\frac{3}{2}} u.
\end{equation}
Then a limit for the 3-Novikov system may be obtained in the limit $\epsilon \rightarrow 0$, that is
\begin{eqnarray}\label{3lim}
\begin{array}{l}u_t+(upr)_x=0,\\
v_t+3vp_xr+v_xpr+u^2p=0,\\
w_t+3wpr_x+w_xpr-u^2r=0,\\
v=-p_{xx}, \quad \quad \quad \quad  w=-r_{xx}.\end{array}
\end{eqnarray}The short wave model (\ref{3lim}) is also integrable in the sense of admitting bi-Hamiltonian structure and a Lax pair. The bi-Hamiltonian structure can be obtained by applying the transformation (\ref{limt}) to that of the 3-Novikov system, that is
\begin{equation}
\left(
  \begin{array}{c}
    u \\
    v \\
    w \\
  \end{array}
\right)_t=\bar{{\cal J}}_1\left(
  \begin{array}{c}
    \frac{\delta \bar{H}_2}{\delta u} \\
    \frac{\delta \bar{H}_2}{\delta v} \\
    \frac{\delta \bar{H}_2}{\delta w} \\
  \end{array}
\right)=\bar{{\cal K}}_1\left(
  \begin{array}{c}
    \frac{\delta \bar{H}_1}{\delta u} \\
    \frac{\delta \bar{H}_1}{\delta v} \\
    \frac{\delta \bar{H}_1}{\delta w} \\
  \end{array}
\right),
\end{equation}where
\begin{eqnarray*}
&&\bar{{\cal J}}_1=\left(
           \begin{array}{ccc}
             \frac{1}{2}\partial  & 0 & 0 \\
             0 & 0 & -\partial^2 \\
             0 & \partial^2 & 0 \\
           \end{array}
         \right),\\
&&\bar{{\cal K}}_1=\left(
                         \begin{array}{ccc}
                           0 & 0 & 0 \\
                           0 & \frac{3}{2}v\partial^{-1}v & -u^2-\frac{3}{2}v\partial^{-1}w \\
                           0 & u^2-\frac{3}{2}w\partial^{-1}v & \frac{3}{2}w\partial^{-1}w \\
                         \end{array}
                       \right)-2\Omega\partial^{-3}\Omega^{*},
\end{eqnarray*}with the functionals given by
\begin{eqnarray*}
\bar{H}_1=\int p_xr_{x} dx,\\
\bar{H}_2=\int (pp_xrr_{xx}+pp_xr_x^2-u^2pr) dx.
\end{eqnarray*}
Taking the transformation (\ref{limt}) to the Lax pair (\ref{Laxpair}) with $\lambda\rightarrow \epsilon \lambda$, a Lax pair for the
limit system (\ref{3lim}) is obtained as
\begin{eqnarray*}
\varphi_x=\left(
            \begin{array}{ccc}
              0 & 1 & 0 \\
              \lambda u^2 & 0 & v \\
              \lambda w & 0 & 0\\
            \end{array}
          \right)
\varphi, \\
\varphi_t=\left(
            \begin{array}{ccc}
              \frac{1}{3\lambda}+pr_x & -pr & \frac{p}{\lambda} \\
              p_xr_x-\lambda u^2pr & \frac{1}{3\lambda}-p_xr & \frac{p_x}{\lambda}-vpr \\
              -\lambda wpr-r_x & r & p_xr-pr_x-\frac{2}{3\lambda}\\
            \end{array}
          \right)
\varphi.
\end{eqnarray*}
Moreover, the short wave model (\ref{3lim}) is also reciprocal connected to the first negative flow in the mYJ hierarchy by taking the similar process before. It  may reduce to that of the Geng-Xue, the Novikov and the DP as $u=0$ and $u=0,p=r$ as well as  $u=0,r=1$ respectively.

\bigskip
\noindent
{\bf Acknowledgments}

This work is partially supported by the National Natural Science Foundation of China (Grant Nos. 11805071 and 11505064) and Promotion Program for Young and Middle-aged Teacher in Science and Technology Research of Huaqiao University (Project No. ZQN-PY301).


\section*{References}
\begin{thebibliography}{20}
\bibitem{Holm}R. Camassa and D. D. Holm, An integrable shallow water equation with peaked solitons, Phys. Rev. Lett. {\bf 71} (1993) 1661-1664.
\bibitem{Lun} A. N. W. Hone, H. Lundmark and  J. Szmigielski, Explicit multipeakon solutions of Novikov's cubically nonlinear integrable Camassa-Holm type equation, Dyn. PDE {\bf 6} (2009) 253-289.
\bibitem{Hyman}R. Camassa, D. D. Holm and J. M. Hyman, A new integrable shallow water equation, Adv. Appl. Mech. {\bf 31} (1994) 1-33.
\bibitem{Fuch}B. Fuchssteiner, Some tricks from the symmetry-toolbox for nonlinear equations: generalizations of the Casmassa-Holm equation, Phys. D {\bf 95} (1996) 229-243.
\bibitem{Lenells}J. Lenells, The correspondence between KdV and Camassa-Holm, IMRN  {\bf 71} (2004) 3797-3811.
\bibitem{Mckean}H. P. McKean, The Liouville correspondence between the Korteweg-de Vries and the Camassa-Holm hierarchies,  Commun. Pure Appl. Math. {\bf 56} (2003) 998-1015.
\bibitem{Qu}J. Song, C. Qu and Z. Qiao, A new integrable two-component system with cubic nonlinearity, J. Math. Phys. {\bf 52} (2011) 013503.
\bibitem{Degas}A. Degasperis and M. Procesi, Asymptotic integrability, in Symmetry and Perturbation Theory edited by Degasperis A and Gaeta G World Scientific (1999) pp23-37.
\bibitem{DHH}A. Degasperis, D. D. Holm and A. N. W. Hone, A new integrable equation with peakon solutions, Theor. Math. Phys. {\bf 133} (2002) 1463-1474.
\bibitem{Novikov}V. S. Novikov, Generalisations of the Camassa-Holm equation, J. Phys. A: Math. Theor. {\bf 42} (2009) 342002.
\bibitem{Wang}A. N. W. Hone  and J. P. Wang, Prolongation algebras and Hamiltonian operators for peakon equations, Inverse Problem {\bf 19} (2003) 129-145.
\bibitem{Geng}X. Geng and B. Xue,  An extension of integrable peakon equations with cubic nonlinearity, Nonlinearity {\bf 22} (2009) 1847-1856.
\bibitem{Hone}A. N. W. Hone and J. P. Wang,  Integrable peakon equations with cubic nonlinearity, J. Phys. A: Math. Theor. {\bf 41} (2008) 372002.
\bibitem{Olver}P. J. Olver and P. Rosenau, Tri-Hamiltonian duality between solitons and solitary-waves solutions having compact support, Phys. Rev. E {\bf 53} (1996) 1900-1906.
\bibitem{Chen}
M. Chen, S. Liu and Y. Zhang, A two-component generalization of the Camassa-Holm equation and its solutions, Lett. Math. Phys. {\bf 75} (2006) 1-15.
\bibitem{Xue} X. Geng and B. Xue, A three-component generalization of Camassa-Holm equation with N-peakon solutions, Adv. Math. {\bf 226} (2011) 827-839.
\bibitem{Li} N. Li and Q. P. Liu, Bi-Hamiltonian structure of a three-component Camassa-Holm type equation, J. Nonlinear
Math. Phys. {\bf 20} (2013) 126-134.
\bibitem{Popowicz} N. Li, Q. P. Liu and Z. Popowicz, A four-component Camassa-Holm type hierarchy, J. Geom. Phys. {\bf 85} (2014) 29-39.
\bibitem{Yajima}N. Yajima and M. Oikawa, Formation and interaction of sonic-Langmuir solitons, Prog. Theor. Phys. {\bf 56} (1976)  1719-1739.
\bibitem{Liu} Q. P. Liu, Modifications of $k$-constrained KP hierarchy, Phys. Lett. A {\bf 187} (1994) 373-381.
\bibitem{Li2}N. Li, Reciprocal link for a three-component Camassa-Holm type equation, J. Nonlinear Math. Phys. {\bf 23} (2016) 150-156.
\bibitem{Lene}J. Lenells, Conservation laws of the Camassa-Holm equation, J. Phys. A: Math. Gen. {\bf 38} (2005)  869-880.
\bibitem{Kang}J. Kang, X. Liu, P. J. Olver and C. Qu, Liouville correspondence between the modified KdV hierarchy and its dual integrable hierarchy, J Nonlinear Sci. {\bf 26} (2016) 141-170.
\bibitem{Pjo}P. J. Olver, Applications of Lie Groups to Differential Equations, 2nd ed., Springer, Berlin, 1993.
\bibitem{Gilson}C. Gilson and A. Pickering, Factorization and Painlev\'e analysis of a class of nonlinear third-order partial differential equations, J. Phys. A: Math. Gen. {\bf 28} (1995) 2871-2888.
\bibitem{hone5}A. N. W. Hone, Reciprocal transformations, Painlev\'e property and solutions of energy-dependent Schr\"{o}dinger hierarchies, Phys. Lett. A {\bf 249} (1998) 46-54.
\bibitem{Oevel}W. Oevel and W. Strampp,  Constrained KP hierarchy and bi-Hamiltonian structures, Commun. Math. Phys. {\bf 157} (1993) 51-81.
\bibitem{Kono}B. G. Konopelchenko and W. Oevel, An r-matrix approach to nonstandard classes of interable equations, Publ. RIMS,  Kyoto Univ. {\bf 29} (1993) 581-666.
\bibitem{Brunelli}J. C. Brunelli and S. Sakovich,  Hamiltonian structures for the Ostrovsky-Vakhnenko equation, Commun. Nonlin. Sci. Numer. Simul. {\bf 18} (2013) 56-62.
\bibitem{Hunt}J. Hunter and R. Saxton, Dynamics of director fields, SIAM J. Appl. Math. {\bf 51} (1991) 1498-1521.
\bibitem{Dai} H. H. Dai and M. Pavlov, Transformations for the Camassa-Holm equation, its high-frequency limit and the Sinh-Gordon equation, J. Phys. Soc.  Japan {\bf 67} (1998) 3655-3657.
\bibitem{Vak}V. A. Vakhnenko, Solitons in a nonlinear model medium, J. Phys. A: Math. Gen. {\bf 25} (1992) 4181-4187.
\end{thebibliography}
\end{document}